\documentstyle[12pt]{article}

\begin{document}
\newcommand {\be}{\begin{equation}}
\newcommand {\ee}{\end{equation}}
\newcommand {\bea}{\begin{array}}
\newcommand {\cl}{\centerline}
\newcommand {\eea}{\end{array}}
\renewcommand {\thefootnote}{\fnsymbol{footnote}}
\baselineskip 0.65 cm
\begin{flushright}
IC/99/179\\
hep-th/9911203
\end{flushright}
\begin{center}
{\Large{\bf A Note on T-Duality, Open Strings in B-field Background and
Canonical Transformations }}
\vskip .5cm

 M.M. Sheikh-Jabbari
\footnote{ E-mail:jabbari@ictp.trieste.it} \\

\vskip .5cm

 {\it The Abdus Salam International Centre for Theoretical Physics 

 Strada Costiera, 11

I-34014, Trieste, Italy}\\
\end{center}

\vskip 2cm
\begin{abstract}
In this paper we study T-duality for open strings ending on branes with non-zero 
B-field on them from the point of view of canonical transformations. For the
particular
case of type II strings on the two torus we show that the $Sl(2,Z)_N$
transformations can be understood as a sub-class of canonical transformations 
on the open strings in the B-field background.

\end{abstract}
\newpage
{\it Introduction}
\newline
Recently noncommutative geometry has been shown to be relevant to D-branes in the
constant B-field background \cite{{CDS},{HD},{AAS},{HV},{SW}}. 
In \cite{CDS}, it was conjectured that a light cone formulation of M-theory with the
non-zero constant 3-from background on a three torus is given by the SYM on
noncommutative two torus, $T^{2}_{\theta}$, and,  considering this background is
crucial to recover the U-duality group of M-theory on $T^3$, $SL(3,Z)\times
Sl(2,Z)_N$, where the deformation parameter of the noncommutative torus , $\theta$,
plays the role of $Sl(2,Z)_N$ metric in the zero volume limit.

In a string theoretic discussion it was shown that quantizing the open strings with
the mixed boundary conditions, the coordinates of end points of such
open strings, $x_i$, which live on a D-brane with 
a non-zero B-field background are really
noncommuting \cite{{AAS},{CHo},{Dir}} and:
\be
[x_i,x_j]=2i\alpha' \left( B(1-B^2)^{-1}\right)_{ij}.
\ee

Besides the noncommutative structure, it was shown that 
the T-duality group of type II theories on $T^2$, $SO(2,2;Z)$ can be
written as $Sl(2,Z)\times Sl(2,Z)_N$, where the first is the mapping class group of
the torus and the second acts on the Kahler moduli of the torus \cite{{AAS},{HV}}.

On the other hand there have been many suggestions in literature trying to explain
T-duality as 
particular canonical transformations \cite{{GIV},{Otto},{YLoz}}. 
This idea was initially studied for the closed strings in general
backgrounds and was extended to the open strings with Dirichlet
boundaries, D-branes. However the case of D-brane bound states, i.e. , open strings
ending on the branes with non-zero B-field \cite{AS} has not been
elaborated on well. 

In this note we will study the behaviour of open strings with different
boundary conditions under the proper canonical transformations and show that
the $Sl(2,Z)_N$ part of T-duality is really a sub-class of these canonical
transformations.

{\it Open string in B-field backgrounds}
\newline
We start with the usual $\sigma$-model action describing the open
strings in B-field backgrounds: 

\be\bea{cc}
S= {1 \over 4\pi\alpha'} \int_{\Sigma} d^2\sigma \bigl[ \eta_{\mu\nu}
\partial_aX^{\mu}\partial_bX^{\nu}g^{ab}+ \epsilon^{ab} B_{\mu\nu}\partial_a
X^{\mu}\partial_bX^{\nu}+ 
{1 \over 2\pi\alpha'}\oint_{\partial \Sigma} d \tau A_i \partial_{\tau}\zeta^i, 
\eea\ee
where $A_{\mu},$ is the $U(1)$ gauge field living on the D-brane and
$\zeta^i$ its internal coordinates. 
The action is invariant under the combined gauge transformation [2]
\be \bea{cc}
B_{\mu\nu}\rightarrow B_{\mu\nu}+\partial_{\mu}\Lambda_{\nu}-\partial_{\nu}\Lambda_{\mu} \\
A_{\mu} \rightarrow A_{\mu}-\Lambda_{\mu}.
\eea\ee
The gauge invariant field strength is then 
\be
{\cal F}_{\mu\nu}=B_{\mu\nu}-F_{\mu\nu} \;\;\;\;\; ,\;\; 
F_{\mu\nu}=\partial_{[\mu}A_{\nu]}.
\ee
So we can always choose $\Lambda_{\mu}$ so that ${\cal
F}_{\mu\nu}=B_{\mu\nu}$, and hereafter we will work in this gauge. 

Variation of the action $S$, leads to the following mixed boundary conditions 

\be
\partial_{\sigma}X^{\mu}+{\cal F}^{\mu}_{\nu} \partial_{\tau}X^{\nu}=0. 
\ee

or in terms of the conjugate canonical momenta, $P_{\mu}$:
  
\be
\partial_{\sigma}X^{\mu}+\alpha' {f}^{\mu\nu}P_{\nu}=0,
\ee
with
\be
f_{\mu\nu}=({\cal F}(1-{\cal F}^2)^{-1})_{\mu\nu}.
\ee
Since we will mostly consider the compactifications of $X^1$ and $X^2$
directions  on two torus, for simplicity we assume that all the ${\cal F}$
components except ${\cal F}_{12}=F$ to be zero and in this case
\footnote{If we  consider the torus radii, $R_1$ and $R_2$, we
find $f={F\over V^2+F^2}$, where $V=R_1R_2$.}  
$$
f_{\mu\nu}=\epsilon_{\mu\nu}f,\;\;\;\;\;\ f={F\over 1+F^2}.
$$
$f$ is the deformation parameter of the noncommutative tours.

Let us return to the question of canonical transformations .
It is well known that, although the canonical transformations 
do not change the equations of motion, they alter the boundary conditions
and hence in our case we expect the boundary conditions, or equivalently
the parameter $f$, to be changed with canonical transformations. But, we
will momentarily show that  for any linear canonical transformation one
can find a transformed $f$, so that the form of the boundary conditions
remain invariant.
This argument can be justified as follows: Boundary conditions can
be treated as the constraints which are of second class \cite{Shir} and
hence the canonical transformations are those under which the {\it Dirac
brackets} are invariant. The above requirement is satisfied if we also 
transform $f$ in a proper way.

Since we are interested in the two torus case, we consider the canonical
transformations which reproduce the $Sl(2,Z)_N$

{\it a)}

\be\left\{\bea{cc}
X_i\rightarrow {\tilde X}_i=\alpha'\int^{\sigma} P_i d\sigma \\
P_i\rightarrow {\tilde P}_i=\partial_{\sigma} X_i,\;\;\;\; i=1,2\; ,
\eea\right.
\ee 
together with \footnote{ For the case of the torus we should take $\alpha'$ to
$\alpha' R_i^2$ and the metric is changed as the usual T-dualities.}   

\be
f\rightarrow {\tilde f}={-1\over f}.
\ee
One can easily show that the above transformations do not change the form
of equations of motion and the boundary conditions,
$$
\partial_{\sigma}X^{\mu}+\alpha' {f}^{\mu\nu}P_{\nu}=0\rightarrow 
\partial_{\sigma}{\tilde X}^{\mu}+\alpha' {\tilde f}^{\mu\nu}{\tilde P}_{\nu}=0.
$$  
In other words  under the above transformations the Dirac bracket introduced in
\cite{{Dir},{Shir}} are invariant. 
It is worth noting that the above transformations are exactly the same as the
transformations introduced in \cite{GIV} for the closed strings.  

{\it b)}

\be\left\{\bea{cc}
X_i\rightarrow {\tilde X}_i=X_i+\epsilon_{ij}\alpha'\int^{\sigma} P_j d\sigma \\
P_i\rightarrow {\tilde P}_i=P_i,\;\;\;\; i=1,2 \; ,
\eea\right.
\ee 
together with  the proper metric transformations (the usual T-duality
transformations), and the $f\rightarrow {\tilde f}={f}+1$, are also the symmetry of
the open string action in the same sense as the case {\it a)}.

As we see the $Sl(2,Z)_N$ transformations which act on the Kahler structure of the 
torus, $\rho=iV+f$, can be generated by the transformations on canonical
variables and also the $f$, background field.
All the above arguments can easily be generalized to the case of $T^n$.

{\it Concluding Remarks}
\newline
In this note we showed that the idea of treating T-duality group of the perturbative
closed string theories (type II strings) as canonical transformations of the
$\sigma$-model can be extended to the strong couplings, using the fact that at
strong coupling type II theories are perturbatively governed by  the open strings 
ending on the D-branes and their bound states.
In this case although the open string boundary conditions (or equivalently the brane
structure we have) change under canonical transformations, we argued that by a 
proper change of parameters we can make our theory to be invariant under such
transformations, which exactly lead to the $Sl(2,Z)_N$ part of the expected
T-duality group. On the other hand it has been argued that this duality can be
realized as the Morita equivalence of the noncommutative gauge bundles governing the
low energy effective theory of the open strings we considered here. So it is an
interesting question to study Morita equivalence in the light of canonical
transformations.

One can address the same procedure in the path integral formulation too. In the path
integral, canonical transformations are translated to changing the operators
by similarity transformations. This point of view has
recently been used for quantizing the discretized M2-branes in a three from field
background \cite{Liho}. In this picture T (or U)-duality is given by a special class
of the unitary transformations which are characterized by C-field. This question
from the string theory point of view can be understood through the DVV
\cite{DVV} string matrix theory.

{\bf Acknowledgements}

I would like to thank H. Arfaei and F. Ardalan for many  fruitful discussions.


\begin{thebibliography}{99}
\bibitem{CDS}

 A. Connes, M.R. Douglas, A. Schwarz,
"Noncommutative Geometry and Matrix Theory: Compactification on Tori",
{\it JHEP} 9802 (1998) 003, hep-th/9711162.

\bibitem{HD} M. R. Douglas, C. Hull, "D-branes and Noncommutative Torus",
{\it JHEP} {\bf 9802} (1998) 008, hep-th/9711165.

Y.-K. E. Cheung, M. Krogh,
"Noncommutative Geometry From $0$-Branes in a Background $B$ Field",
hep-th/9803031.

F. Ardalan, H. Arfaei, M. M. Sheikh-Jabbari,
"Mixed Branes and M(atrix) Theory on Noncommutative Torus", hep-th/9803067.

\bibitem{AAS} 
F. Ardalan, H. Arfaei, M. M. Sheikh-Jabbari,
"Noncommutative Geometry form Strings and Branes", 
{\it JHEP} {\bf 02} (1999) 016, hep-th/9810072.

\bibitem{HV}
C. Hofman and E. Verlinde, "U-duality of Born-Infeld on the Noncommutative  
Two Torus", 
{\it JHEP} {\bf 9812} (1998) 010, hep-th/9810116.
	
\bibitem{SW}
N. Seiberg, E. Witten, "String Theory and Noncommutative Geometry", 
{\it JHEP} {\bf 09} (1999) 032, hep-th/9908142.



\bibitem{CHo} C-S. Chu and P-M. Ho, "Noncommutative Open Strings and D-branes",
{\it Nucl. Phys.} {\bf B550} (1999) 151, hep-th/9812219.

\bibitem{Dir}
F. Ardalan, H. Arfaei, M. M. Sheikh-Jabbari, "Dirac Quantization of Open Strings and
and Noncommutativity in Branes", hep-th/9906161.

C-S. Chu, P.M. Ho, "Constrained Quantization of Open Strings in Background B-Field
and Noncommutative D-Branes", hep-th/9906192.  

\bibitem{GIV}
A. Giveon, E. Rabinovici and G. Veneziano, {\it Nucl. Phys.} {\bf B322} (1989) 167.

E. Alvarez, L. Alvarez-Gaume' and Y. Lozano, 
{\it Phys. Lett.} {\bf B336} (1994) 183.

A. Giveon, M. Poratti and E. Rabinovici, {\it Phys. Rep.} {\bf 244} (1994) 77.

\bibitem{Otto}

H. Dorn, H.-J. Otto, {\it Phys. Lett.} {\bf B381} (1996) 81.

\bibitem{YLoz}

Y. Lozano, {\it Mod. Phys. Lett.} {\bf A11} (1996) 2893, hep-th/9610024 and
references therein.

\bibitem{AS}
E. Gava, K.S. Narain, M.H. Sarmadi, "On the Bound State of p- and (p+2)-Branes",
{\it Nucl. Phys.} {\bf B504} (1997) 214, hep-th/9704006.

H. Arfaei, M.M. Sheikh-Jabbari, "Mixed Boundary Conditions and Brane-String Bound
States", {\it Nucl. Phys.} {\bf B526} (1998) 278, hep-th/9709054.

M.M. Sheikh-Jabbari, "More on Mixed Boundary Conditions and D-brane Bound States",
{\it Phys. Lett.} {\bf B425} (1998) 48.  

\bibitem{Shir}
M.M. Sheikh-Jabbari and A. Shirzad, "Boundary Conditions as Dirac Constraints ", 
hep-th/9907055.

\bibitem{Liho}
C-S. Chu, P-M. Ho, M. Li, "Matrix Theory in a Constant C Field Background",
hep-th/9911153.

\bibitem{DVV}
R. Dijgraaf, E. Verlinde, H. Verlinde, "Matrix String Theory", 
{\it Nucl. Phys.} {\bf B500} (1997) 43, hep-th/9703030.

\end{thebibliography}
\end{document}